\documentclass[aps,prl,superscriptaddress,twocolumn,showpacs]{revtex4}
\usepackage{bm}
\usepackage{graphicx}
\usepackage{amsfonts}
\usepackage{amsmath}

\begin{document}

\title{Quantum 2D Heisenberg antiferromagnet:\\ bridging the gap
       between field-theoretical and semiclassical approaches}

\author{Bernard~B. Beard}
\affiliation{Departments of Physics and Mechanical Engineering of the
             Christian Brothers University
             - Memphis, Tennessee 38104, USA}
\author{Alessandro Cuccoli}
\affiliation{Dipartimento di Fisica dell'Universit\`a di Firenze -
             via G. Sansone 1, I-50019 Sesto Fiorentino (FI), Italy}
\affiliation{Istituto Nazionale per la Fisica della Materia
             - U.d.R. Firenze -
             via G. Sansone 1, I-50019 Sesto Fiorentino (FI), Italy}
\author{Ruggero Vaia}
\affiliation{Istituto di Fisica Applicata `Nello Carrara'
             del Consiglio Nazionale delle Ricerche\,-\,via
             Panciatichi~56/30, I-50127 Firenze, Italy}
\affiliation{Istituto Nazionale per la Fisica della Materia
             - U.d.R. Firenze -
             via G. Sansone 1, I-50019 Sesto Fiorentino (FI), Italy}
\author{Paola Verrucchi}
\affiliation{Dipartimento di Fisica dell'Universit\`a di Firenze -
             via G. Sansone 1, I-50019 Sesto Fiorentino (FI), Italy}
\affiliation{Istituto Nazionale per la Fisica della Materia
             - U.d.R. Firenze -
             via G. Sansone 1, I-50019 Sesto Fiorentino (FI), Italy}
\date{\today}

\begin{abstract}
The field-theoretical result for the low-$T$ behaviour of the
correlation length of the quantum Heisenberg antiferromagnet on the
square lattice was recently improved by Hasenfratz [Eur.~Phys.~J. B
{\bf 13},~11 (2000)], who corrected for cutoff effects. We show that
starting from his expression, and exploiting our knowledge of the
classical thermodynamics of the model, it is possible to take into
account non-linear effects which are responsible for the main features
of the correlation length at intermediate temperature. Moreover, we
find that cutoff effects lead to the appearance of an effective
exchange integral depending on the very same renormalization
coefficients derived in the framework of the semiclassical {\em
pure-quantum self-consistent harmonic approximation}: The gap between
quantum field-theoretical and semiclassical results is here eventually
bridged.
\end{abstract}


\maketitle

In the last two decades, thermodynamic properties of the quantum
Heisenberg antiferromagnet on the square-lattice (QHAF) have been
determined by a number of substantially different methods. Theoretical
predictions were compared with experimental data for real
compounds~\cite{Expt}, as well as with numerical simulations obtained
by different methods~\cite{BBB,KimT1998}, and also with high-$T$ series
expansions~\cite{ElstnerSSGB1995}. Despite this effort, however, a
comprehensive picture of the subject has not yet been formulated.

Much of the analysis and debate on the QHAF has focused on the
temperature- and spin-dependence of the staggered correlation length
$\xi(T,S)$. Goal of this paper is to show that $\xi(T,S)$ can be
expressed as
\begin{equation}
 \xi(T,S) = \xi_{\rm{cl}}\bigg(\frac T{J_{\rm{eff}}(T,S)}\bigg)~,
\label{e.xiTeff}
\end{equation}
where the effective exchange integral $J_{\rm{eff}}(T,S)$ embodies
quantum effects and is defined in terms of purely quantum
spin-fluctuations, while $\xi_{\rm{cl}}(T/J_{\rm{cl}})$ is the
correlation length of the classical HAF. In particular, we find that
Eq.~\eqref{e.xiTeff}, besides being the outcome of the semiclassical
{\em pure-quantum self-consistent harmonic approximation}
(PQSCHA)~\cite{CTVV199697}, remarkably holds also for the quantum
field-theoretical prediction~\cite{QFT}, as recently improved by
Hasenfratz~\cite{Hasenfratz2000}. On the other hand, we show that the
PQSCHA, when properly designed to such purpose, allows for a correct
description of the low-$T$ regime, {\it via} the appearance of the very
same $J_{\rm{eff}}(T,S)$ implicitly entering Hasenfratz's expression.

It is a definite suprise that a typical semiclassical expression such
as Eq.~\eqref{e.xiTeff}, come out of a field-theoretical result,
especially in the case of the QHAF, where semiclassical and
field-theoretical approaches seemed destined to describe different
regions of the $(T,S)$ plane, with no possible overlap.

The QHAF is defined by the spin Hamiltonian
\begin{equation}
 \hat{\cal{H}} = \frac J2 \sum_{{\bm{i}}{\bm{d}}}
 \hat{{\bm{S}}}_{\bf i}\cdot \hat{{\bm{S}}}_{{\bm{i}}+{\bm{d}}}~,
\label{e.QHAF}
\end{equation}
where $J{>}0$, ${\bm{i}}{=}(i_1,i_2)$, ${\bm{d}}{=}(\pm{1},\pm{1})$,
and lengths are in lattice units; the spin operators
$\hat{\bm{S}}_{\bm{i}}\,{=}\,
(\hat{S}_{\bm{i}}^x,\hat{S}_{\bm{i}}^y,\hat{S}_{\bm{i}}^z)$ satisfy
$|\hat{\bm{S}}_{\bm{i}}|^2{=}S(S{+}1)$.

The correlation length $\xi(T,S)$ is defined {\em{via}} the asymptotic
behaviour,
$\lim_{|{\bm{r}}|\to\infty}G({\bm{r}})\propto{e}^{-|{\bm{r}}|/\xi}$, of
the staggered correlation function $G({\bm{r}}){\equiv}(-1)^{r_1+r_2}
\langle\hat{S}_{\bm{i}}^z\hat{S}_{{\bm{i}}+{\bm{r}}}^z\rangle$~,
with ${\bm{r}}{=}(r_1,r_2)$ any vector on the square lattice.

The classical counterpart of the QHAF corresponds to the limit
$S{\to}\infty$ with $JS^2{\to}{J_{\rm{cl}}}$, that gives the classical
Hamiltonian
 ${\cal{H}}{=}(J_{\rm{cl}}/2)\sum_{{\bm{i}}{\bm{d}}}
 {\bm{s}}_{\bm{i}}{\cdot}{\bm{s}}_{{\bm{i}}+{\bm{d}}}~$,
where ${\bm{s}}_{\bm{i}}$ are classical unit vectors and $J_{\rm{cl}}$
is the only energy scale of the model.

Let us consider the right-hand side of Eq.~\eqref{e.xiTeff}: The 3-loop
analytical expression for the classical correlation length, as from the
field-theoretical approach~\cite{CFT}, is
\begin{equation}
 \xi_{\rm{cl(3l)}}\Big(\!\frac{T}{J_{\rm{cl}}}\Big) =
 \frac{e^{1-\pi/2}}{8\sqrt{32}} \frac{T}{2\pi J_{\rm{cl}}}
 \,e^{\textstyle\frac{2\pi J_{\rm{cl}}}T}
 \Big[1{-}\frac{0.574~T}{2\pi J_{\rm{cl}}} {+}O(T^2)\Big].
\label{e.xicl}
\end{equation}
This expression is quantitatively correct up to
$T/J_{\rm{cl}}\lesssim{0.4}$~. Above this temperature it smoothly
connects to the exact Monte Carlo data for the classical
model~\cite{CMC}, so that the exact classical correlation length
$\xi_{\rm{cl}}(T/J_{\rm{cl}})$ gets available for all temperatures.

As for the left-hand side of Eq.~\eqref{e.xiTeff}, the most celebrated
field-theoretical result is the 3-loop expression derived from the
correlation length of the quantum nonlinear $\sigma$-model
(QNL$\sigma$M)~\cite{QFT},
\begin{equation}
 \xi_{\rm{3l}}(T,S) = \frac e8 \frac c{2\pi{\rho_{{}_S}}}
 ~e^{\textstyle\frac{2\pi{\rho_{{}_S}}}T}
 ~\Big[1{-}\frac12\frac T{2\pi{\rho_{{}_S}}}
 {+}O\Big(\frac{T^2}{S^4}\Big)\Big]~.
\label{e.xi3l}
\end{equation}
The mapping to the QHAF consists of identifying $c$ and ${\rho_{{}_S}}$
with the zero-$T$ renormalized spin velocity and spin stiffness of the
system
\begin{equation}
 c=2\sqrt{2}JSZ_c(S)~,~~~{\rho_{{}_S}}=JS^2Z_\rho(S)~,
\label{e.crho}
\end{equation}
with $Z_c(S)$ and $Z_\rho(S)$ quantum renormalization coefficients; for
$c$ and ${\rho_{{}_S}}$ we will hereafter use the most accurate
available values~\cite{SWT}, as from their expansion up to $O(S^{-2})$
and $O(S^{-3})$, respectively.

One of the essential features of Eq.~\eqref{e.xi3l} is the
temper\-ature-independent pre-exponential factor
$(e/8)(c/2\pi{\rho_{{}_S}})$, which contrasts with the $O(T)$ prefactor
of the classical Eq.~\eqref{e.xicl}. The $O(1)$ prefactor gives a
purely exponential asymptotic behaviour $\xi\sim{e^{A/T}}$, with
$A\sim{2\pi{J}S^2}$, which is also the outcome of other theoretical
methods such as the Schwinger-boson approach~\cite{ArovasA1988} and the
modified spin-wave theory~\cite{Takahashi1989}. Thus the behaviour
$\xi\sim{e^{A/T}}$ has been generally regarded as a signature of the
quantum character of the model. Early numerical studies suggested this
signature is observed in the experimentally accessible temperature
region of $S=1/2$ antiferromagnets~\cite{Mak91}. In the last few years,
however, this assumption has proved misleading in subtle ways, as
severe difficulties arose when approaching the $S{\ge}1$ case by the
effective field-theory. Finally, in Ref.~\onlinecite{BBB} it was
clearly shown that Eq.~\eqref{e.xi3l} holds only for temperatures low
enough to ensure an extremely large correlation length, e.g.,
$\xi\gtrsim{10^5}$ for $S=1$, $\xi\gtrsim{10^{12}}$ for $S=3/2$, and
generally cosmological correlation lengths for $S>3/2$.

In Ref.~\onlinecite{Hasenfratz2000} Hasenfratz explained why cutoff
effects, which are so devious for $S=1/2$, significantly modify the
correlation length for $S\ge{1}$. By exploiting a direct mapping
between the QHAF and the QNL$\sigma$M, Hasenfratz obtained the
cutoff-corrected field-theoretical result
\begin{equation}
 \xi_{{}_{\rm{H}}}(T,S) = \xi_{\rm{3l}}(T,S)~e^{-C(T,S)}~,
\label{e.xiH}
\end{equation}
where $C(T,S)$, defined in Eq.~(14) of
Ref.~\onlinecite{Hasenfratz2000}, is an integral of familiar spin-wave
quantities over the first Brillouin zone. With this correction, which
is the leading order in the spin-wave expansion for the cutoff
correction, it is possible to obtain numerically accurate results down
to $\xi\gtrsim 10^3$ for all $S$. Eq.~\eqref{e.xiH} is therefore the
best available prediction of the field-theoretical approach.

Our first step is to note that the function $C(T,S)$ may be written as
\begin{eqnarray}
 && C(T,S) \equiv {\frac\pi 2} + \ln\frac{16JS\zeta_1(0,S)}T~
\nonumber\\
 &&\hspace{12mm}
 -\frac{2\pi JS^2\zeta_1(0,S)}T
 \big[\delta\zeta_1(T,S)+\delta\zeta_0(T,S)\big]~,~~~~~
\label{e.C}
\end{eqnarray}
where $\delta\zeta_i(T,S)=\zeta_i(T,S)-\zeta_i(0,S)=O(T/S^2)$, and
\begin{equation}
 \zeta_i(T,S)=1{+}\frac1{2S}{-}\frac1{2SN}\sum_{\bm{k}}
 \frac{(1{-}\gamma_{\bm{k}}^2)^{1/2}}{(1{-}\gamma_{\bm{k}})^{1{-}i}}
 ~{\cal{L}}_{\bm{k}}(T,S),
\label{e.zetai}
\end{equation}
\begin{eqnarray}
 {\cal{L}}_{\bm{k}}(T,S)&=&
 \coth \frac{\omega_{\bm{k}}(S)}{2\,T}-\frac{2\,T}{\omega_{\bm{k}}(S)}~,
\label{e.Lk}\\
 \omega_{\bm{k}}(S) &=& 4JS\zeta_1(0,S)~\sqrt{1-\gamma_{\bm{k}}^2}~,
\label{e.omegak}
\end{eqnarray}
with sums over wavevectors ${\bm{k}}=(k_1,k_2)$ in the first Brillouin
zone, and $\gamma_{\bm{k}}=(\cos k_1+\cos k_2)/2~$.

From the above formulas we find
\begin{equation}
 \xi_{{}_{\rm{H}}}(T\!,S) = \alpha_1(T\!,S)
 \,\xi_{\rm{cl(3l)}}\Big(\frac{T}{JS^2}\Big)
 \,\alpha_2(T\!,S)\,\alpha_3(T\!,S)\,,
\label{e.xiHcl}
\end{equation}
with
\begin{eqnarray*}
 \alpha_1(T,S)&=&{\textstyle\frac{Z_c}{Z_\rho}}
 ~e^{{\textstyle\frac{2\pi JS^2}T}(Z_\rho-1)}~,
\label{e.a1}\\
 \alpha_2(T,S)&=&{\textstyle\frac1{\zeta_1(0,S)}}
 ~e^{{\textstyle\frac{2\pi JS^2}T}\,\zeta_1(0,S)
 (\delta\zeta_1+\delta\zeta_0)}~,
\label{e.a2}\\
 \alpha_3(T,S)&=&\textstyle\Big[1-\frac{T}{2\pi JS^2}
                \Big(\frac1{2Z_\rho}-\textstyle{0.574}\Big)
                +O\Big(\frac{T^2}{S^4}\Big)\Big]~;
\label{e.a3}
\end{eqnarray*}
notice that $\alpha_1$ is an exact coefficient, while $\alpha_2$ is
given at the first order in $1/S$ as from
Ref.~\onlinecite{Hasenfratz2000}: this is a signature of the different
approaches used in determining $\xi_{\rm{3l}}$ and $C(T,S)$. As for
$\alpha_3$, it contains the 3-loop correction terms of the
field-theoretical results.

In order to single out the relevant temperature scale in the
exponential factor of Eq.~\eqref{e.xiHcl}, one can embody the
$O(S^{-1})$ terms of the leading exponential in the argument of
$\xi_{\rm{cl(3l)}}$: to this end we explicitly write $Z_\rho$ and $Z_c$
from their spin-wave theory expression~\cite{SWT}
\begin{equation}
 Z_c=\zeta_1(0,S)+\Delta Z_c~,~~~
 Z_\rho=\zeta_1(0,S)\zeta_0(0,S)+\Delta Z_\rho~,
\label{e.Zcrho}
\end{equation}
where $\Delta Z_c$ and $\Delta Z_\rho$ are $O(S^{-2})$, and
find
\begin{equation}
 \xi_{{}_{\rm{H}}}(T,S)=
 \xi_{\rm{cl(3l)}}\bigg(\frac T{J_{\rm{eff(H)}}(T,S)}\bigg)
~\alpha(T,S)
\label{e.xiHTeff}
\end{equation}
with
\begin{eqnarray}
  &&~~~J_{\rm{eff(H)}}(T,S) = JS^2~\zeta_0(T,S)~\zeta_1(T,S)~,
\label{e.JeffH}
\\
 &&\alpha(T,S) =
 \big[(1+\delta\zeta_1+\delta\zeta_0 +O(S^{-2})\big]
\nonumber\\
 &&\hspace{13mm}\times~
 \textstyle\Big[1-\frac{T}{2\pi}
 \Big(\frac1{2 JS^2Z_\rho}-\frac{0.574}{J_{\rm{eff(H)}}}\Big)
 +O\Big(\frac{T^2}{S^4}\Big)\Big]\,.~~~~~~
\label{e.appalpha}
\end{eqnarray}

At this level, the classical correlation length enters
Eq.\eqref{e.xiHTeff} by its 3-loop NL$\sigma$M expression,
Eq.\eqref{e.xicl}. However, inspired by the PQSCHA result of
Ref.~\onlinecite{CTVV199697}, we are led to generalize
Eq.~\eqref{e.xiHTeff} by replacing $\xi_{\rm{cl(3l)}}$ with the exact
$\xi_{\rm{cl}}$, which is numerically available at all temperatures.
This is a fundamental step in order to get Eq.~\eqref{e.xiHTeff} to
reproduce the experimental and quantum Monte Carlo (QMC) data in the
intermediate temperature range, a goal that Eq.~\eqref{e.xiH} does not
accomplish yet.

\begin{figure}
\includegraphics[bbllx=14mm,bblly=20mm,bburx=190mm,bbury=255mm,%
     width=63mm,angle=90]{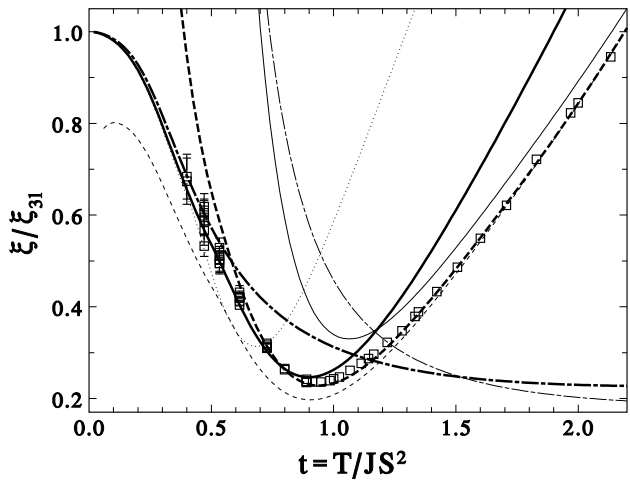}
\caption{\label{f.Fig1}
Ratio $\xi/\xi_{\rm{3l}}$ {\it vs} $T/JS^2$ for $S=5/2$. Dash-dotted
line: Hasenfratz's result~\eqref{e.xiH}; dotted and solid lines:
Eqs.~\eqref{e.xiHcl} and~\eqref{e.xiHTeff} with $\xi_{\rm{cl(3l)}}$
replaced by the exact $\xi_{\rm{cl}}$; thin dashed line:
Eq.~(\ref{e.xiTeff}) with $J_{\rm{eff(H)}}$; dashed line: PQSCHA
result, i.e., Eqs.~\eqref{e.xiTeff} and~\eqref{e.JeffPQ}; thin solid
and dash-dotted lines: $\xi_{\rm{cl}}/\xi_{\rm{3l}}$ and
$\xi_{\rm{cl(3l)}}/\xi_{\rm{3l}}$, respectively (see text); the symbols
are our QMC data.}
\end{figure}

In Fig.~\ref{f.Fig1} we show the ratio $\xi/\xi_{\rm{3l}}$ as a
function of $T/JS^2$ for $S=5/2$. The solid line is obtained by
replacing $\xi_{\rm{cl(3l)}}$ in Eq.(\ref{e.xiHTeff}) with the exact
$\xi_{\rm{cl}}$. Compared with Hasenfratz's expression (dash-dotted
line), the agreement with our QMC data greatly improves, just because
the classical correlation length is accounted for exactly.

It is worthwhile mentioning that one could use the exact
$\xi_{\rm{cl}}$ already in Eq.~(\ref{e.xiHcl}) (dotted line in
Fig.~\ref{f.Fig1}): such curve shows that the pronounced minimum in the
ratio $\xi/\xi_{\rm{3l}}$, a feature that field-theoretical results do
not reproduce, is due to substantially classical nonlinear effects
which need a non-perturbative (or exact) treatment; this is confirmed
by comparing the classical exact $\xi_{\rm{cl}}$ (thin solid) and
3-loop $\xi_{\rm{cl(3l)}}$ (thin dash-dotted) lines, reported in
Fig.~\ref{f.Fig1} by fixing $J_{\rm{cl}}$ to the classical limit
(${\cal{L}}_{\bm{k}}{=}0$) of $J_{\rm{eff(H)}}$. The correct position
of the minimum, on the other hand, is found by properly singling out
the dominant temperature scale for the quantum correlation length,
i.e., by determining the appropriate effective exchange integral. To
this respect, by comparing Eq.~\eqref{e.JeffH} with Eqs.~\eqref{e.crho}
and
\eqref{e.Zcrho}, one can see that  $J_{\rm{eff(H)}}(T,S)$ tends to
${\rho_{{}_S}}$ for $T{\to}{0}$ and can be actually interpreted as a
{\it temperature-dependent} spin stiffness ${\rho_{{}_S}}(T)$.

Finally, we drop the 3-loop correction terms, set the residual
$\alpha(T,S)$ to unity, and eventually obtain Eq.~(\ref{e.xiTeff}),
with the effective exchange integral Eq.~(\ref{e.JeffH}). The
corresponding curve is shown in Fig.~1 by the thin dashed line. The
anomalous behaviour seen as $T\,{\to}\,0$ has no physical meaning: in
fact, when considering $\xi$ as in Eq.~(\ref{e.xiTeff}) with
$J_{\rm{eff(H)}}$, the ratio $\xi/\xi_{\rm{3l}}$ contains a vanishing
factor $\exp(-2\pi{J}S^2\Delta{Z_\rho}/T)$, due to the fact that
${\rho_{{}_S}}$ in $\xi_{\rm{3l}}$ is taken up to $O(S^{-3})$, while
$J_{\rm{eff(H)}}$ is only taken up to $O(S^{-2})$. To this respect, it
is worthwhile noticing that Hasenfratz's expression~\eqref{e.xiH} is
spurious as far as the order in $1/S$ is concerned, because $C(T,S)$
and $\xi_{\rm{3l}}$ are accurate up to $O(S^{-1})$ and $O(S^{-3})$,
respectively. This originates few other inconsistencies in the $1/S$
approximation, essentially contained in the $O(S^{-2})$ terms of
Eq.~\eqref{e.appalpha}, which result in slight discrepancies from the
expected values. Nevertheless, the thin dashed line reproduces the
behaviour drawn by QMC data in the whole temperature range and bridges
the low- and intermediate-$T$ regime, a success never scored before.

Let us now comment on the above result. First of all, we notice that
$J_{\rm{eff(H)}}$ depends just on the {\em pure-quantum} part of the
spin-fluctuations, i.e., the difference between the quantum and the
classical spin-fluctuations~\cite{CGTVV1995} (the signature of this
being the Langevin functions Eq.~\eqref{e.Lk}). In particular, the
renormalization factors $\zeta_0(T,S)$ and $\zeta_1(T,S)$ represent the
effect of the pure-quantum fluctuations of each spin with respect to
its local alignment axis and to its nearest-neighbors, respectively.

By considering Eq.~(\ref{e.JeffH}) with the low-$T$ expansions
\begin{eqnarray*}
 \delta\zeta_0(T,S)&\mathop\simeq\limits_{T\to{0}}&\!
 \frac T{2\pi JS^2\zeta_1(0,S)}\ln \frac{16JS\zeta_1(0,S)}{T}+O(T^3),
\\
 \delta\zeta_1(T,S)&\mathop\simeq\limits_{T\to{0}}&\!
 \frac T{4JS^2\zeta_1(0,S)}+O(T^3)~,
\end{eqnarray*}
one sees that it is the pure-quantum on-site renormalization parameter
$\zeta_0(T,S)$ that cancels the classical-like pre-exponential factor.
This suggests that the asymptotic behaviour of Eq.~\eqref{e.xi3l} sets
in when pure-quantum fluctuations of one spin relative to its nearest
neighbors become negligible, and spins are mainly affected by on-site
fluctuations of pure-quantum origin.

At this point, we recall that the pure-quantum renormalization
coefficients $\zeta_0(T,S)$ and $\zeta_1(T,S)$ are characteristic of
the PQSCHA, by which the correlation length of the QHAF in the form of
Eq.~\eqref{e.xiTeff} was first obtained~\cite{CTVV199697} with
\begin{equation}
 J_{\rm{eff}}(T,S) = JS^2~\zeta_1^2(T,S)~.
\label{e.JeffPQ}
\end{equation}

The resemblance between Eqs.~\eqref{e.JeffH} and~\eqref{e.JeffPQ} is
remarkable, given the fact that the PQSCHA is a completely different
method from that used by Hasenfratz. However, the PQSCHA fails to
describe the regime of very low-$T$, e.g., for $S{=}1/2$ it holds only
for
$T{\gtrsim}{0.2}\,J$~\cite{CTVV199697,PRLcommrepl1997,Hasenfratz2000}.

In particular, the PQSCHA has been criticized because the expression
Eq.~\eqref{e.JeffPQ} does not lead to the correct low-$T$ asymptotic
behaviour of $\xi$. In fact, we here show that the PQSCHA may be
properly designed to describe the low-$T$ regime, thus reproducing
Eq.~\eqref{e.xi3l}, though with $c$ and ${\rho_{{}_S}}$ only given at
the first order in $1/S$. This is an essential issue, as it means that
the QHAF, even in the low-$T$ limit, is a system with separable
classical and pure-quantum aspects.

One of the main steps in deriving the PQSCHA for a magnetic system is
the construction of an effective classical spin Hamiltonian from that
written in terms of canonical variables~\cite{CGTVV1995}. For the sake
of clarity, we use $({\bm{p}},{\bm{q}}){=}\{p_{\bm{i}},q_{\bm{i}}\}$
and ${\bm{s}}{=}\{{\bm{s}}_{\bm{i}}\}$ to indicate the set of classical
canonical variables and unit vectors, respectively, needed to describe
the classical system (see Ref.~\onlinecite{CTVV199697} for detailed
definitions). For the partition function of the QHAF one finds
 ${\cal{Z}}{=}\int d{\bm{p}}
 d{\bm{q}}\exp(-{\cal{H}}({\bm{p}},{\bm{q}})_{\rm eff}/T)$ with
\begin{equation}
 {\cal{H}}_{\rm{eff}}({\bm{p}},{\bm{q}})=
 \theta^4{\cal{H}}\big(
 {\textstyle\frac{{\bm{p}}}{\theta},\frac{{\bm{q}}}{\theta}}\big)
 \rightarrow\theta^4{\cal{H}}({\bm{s}})
\label{e.Heff}
\end{equation}
and $\theta^2{\equiv}(1+1/2S)^{-1}\zeta_1$; the further scaling
${\bm{p}}\to\theta{\bm{p}}$ and ${\bm{q}}\to\theta{\bm{q}}$ reproduces
the classical-like Heisenberg Hamiltonian
 ${\cal{H}}({\bm{s}}){=}J(S{+}1/2)^2\theta^4\sum_{{\bm{i}}{\bm{d}}}
 {\bm{s}}_{\bm{i}}{\cdot}{\bm{s}}_{{\bm{i}}+{\bm{d}}}~$,
thus defining $J_{\rm{eff}}$ as in Eq.~\eqref{e.JeffPQ}. When the
correlation functions are considered, different scaling laws come into
play, {\it via} the integral
\begin{equation}
 G({\bm{r}}) \propto {\textstyle\int} d{\bm{p}}~d{\bm{q}}
 ~g_{\rm{eff}}({\bm{r}};{\bm{p}},{\bm{q}})~
 e^{{-{\cal{H}}_{\rm{eff}}({\bm{p}},{\bm{q}})/T}}
\label{e.Gr}
\end{equation}
where $g_{\rm{eff}}$ scales according to
\begin{equation}
 g_{\rm{eff}}({\bm{r}};{\bm{p}},{\bm{q}})=\theta^4_{\bm{r}}~
 g\big({\bm{r}};{\textstyle\frac{{\bm{p}}}{\theta_{\bm{r}}},
 \frac{{\bm{q}}}{\theta_{\bm{r}}}} \big)
 \rightarrow\theta^4_{\bm{r}}~ g({\bm{r}};{\bm{s}})
\end{equation}
with $g({\bm{r}};{\bm{s}}){=}N^{-1}\sum_{\bm{i}}
{\bm{s}}_i{\cdot}{\bm{s}}_{{\bm{i}}+{\bm{r}}}$; $\theta^2_{\bm{r}}$ is
a pure-quantum renormalization coefficient defined in
Ref.~\onlinecite{CTVV199697}. As both ${\cal{H}}({\bm{p}},{\bm{q}})$
and $g({\bm{r}};{\bm{p}},{\bm{q}})$ are biquadratic functions, the
different scalings (with $\theta$ or $\theta_{\bm{r}}$) conflict. At
intermediate $T$, thermodynamics is governed by many different spin
configurations, and the effective Hamiltonian appearing in the
Boltzmann factor in Eq.~\eqref{e.Gr} dominates (i.e., scaling with
$\theta$ is preferable~\cite{CTVV199697}): this gives
Eq.~\eqref{e.JeffPQ}. At low-$T$ the quartic terms in the effective
Hamiltonian do not matter, and the scaling conflict is settled:
therefore, the best evaluation of the integral~\eqref{e.Gr} is made by
scaling with $\theta_{\bm{r}}$, so that the prefactor $\theta^4$ in
Eq.~\eqref{e.Heff} is replaced by $\theta^2\theta_{\bm{r}}^2$. As for
increasing $|{\bm{r}}|$ the coefficient $\theta_{\bm{r}}$ rapidly
converges to $(1{+}1/2S)^{-1}\zeta_0$, the PQSCHA effective exchange
for the low-$T$ correlation length reads ${JS^2}\,\zeta_0\,\zeta_1$,
consistent with Eq.~\eqref{e.JeffH}.

A quantitative comparison between Hasenfratz's and PQSCHA results
should also take into account some important details concerning the
spin-wave dispersion relation used to evaluate the pure-quantum
renormalization factors Eqs.~(\ref{e.zetai}). We just mention that in
order to avoid ambiguities due to the ordering prescription one should
definitely prefer the zero-$T$ renormalized
frequencies~\eqref{e.omegak} to the bare frequencies originally used by
Hasenfratz. In fact, the PQSCHA method~\cite{CTVV199697,CGTVV1995} uses
even more refined temperature- and configuration-dependent renormalized
frequencies; treating them within the so-called {\em low-coupling
approximation} yields a more accurate finite-$T$ expression, which is
crucial for quantitative accuracy in evaluating thermodynamic
quantities~\cite{CTVV199697}.

In this paper we have shown that, when cutoff effects are properly
taken into account, the field theoretical expression for the
correlation length of the QHAF analitically contains its classical
counterpart, according to a simple equation that gives the former in
terms of the latter with an effective exchange integral. Such equation
surprisingly holds at all temperatures. Once the exact classical
correlation length is available, the above result is used to make clear
that the main features of the quantum correlation length at
intermediate temperatures are due to essentially classical non-linear
effects, which cannot be taken into account by perturbative approaches.
Moreover, the effective exchange integral is seen to depend on the same
pure-quantum renormalization coefficients defined by the PQSCHA,
according to an expression which is very similar (equal) to that found
by the latter approach in its standard (low-$T$) version.

The idea that an effective exchange integral may be uniquely defined
for all thermodynamic quantities of the QHAF, as predicted by the
PQSCHA, is suggestive and deserves, in our opinion, further
investigation by the field-theoretical community.

\begin{acknowledgments}
This work was supported by MURST in the framework of the COFIN00
program No. MM02102319 and by INFM through the IS-Sezione D (2001).
B.~B.~B. wishes to thank the Physics Department of the University of
Firenze for hospitality.
\end{acknowledgments}


\end{document}